# Systematic dispersion compensation for spectral domain optical coherence tomography using time-frequency analysis and iterative optimization for iridocorneal angle imaging

SHANGBANG LUO,[1,2] GUY HOLLAND,[3] ERIC MIKULA,[2,3] SAMANTHA BRADFORD,[2] REZA KHAZAEINEZHAD,[4] JAMES V JESTER,[1,2] AND TIBOR JUHASZ[1,2,3,*]

[1]*Department of Biomedical Engineering, University of California, Irvine, Irvine, CA 92697, USA*
[2]*Department of Ophthalmology, University of California, Irvine, Irvine, CA 92697, USA*
[3]*ViaLase Inc., Aliso Viejo, CA 92656, USA*
[4]*Beckman Laser Institute, University of California, Irvine, Irvine, CA 92697, USA*
*\*tjuhasz@hs.uci.edu*

**Abstract:** Dispersion is a common phenomenon in optics due to the frequency dependence of the refractive index in polychromatic light. This issue, if left untreated in optical coherence tomography (OCT) imaging, leads to signal broadening of the coherence length and deterioration of the axial resolution. We report a new numeric method for the systematic dispersion compensation in a spectral-domain (SD) OCT for imaging the iridocorneal angle of human cadaver eyes. The dispersion compensation for our OCT system is calculated by an automated iterative process that minimizes the wavenumber-dependent variance of the ridge extracted from the energy distribution of a mirror's spectral interferogram using Short-Time Fourier Transform (STFT) Time-Frequency Analysis (TFA). The average axial resolution of 2.7 $\mu m$ in air was achieved at a range of depths up to 2 $mm$. Compensated OCT images of the iridocorneal angle in human cadaver eyes were much clearer than non-compensated images. We demonstrate the feasibility, effectiveness, and robustness of the proposed method for dispersion compensation in an SD-OCT by evaluating both the mirror and human cadaver eye measurements. We also verified that our imaging system is able to visualize the iridocorneal angle details, such as trabecular meshwork (TM), Schlemm's canal (SC), and collector channels (CCs), which are important ocular outflow structures and play a crucial role in glaucoma managements.

## 1. Introduction

Optical coherence tomography (OCT) is a micron-scale, non-contact three-dimensional imaging technique that employs low coherence light to derive the structural information within optically scattering media, such as biological tissues [1]. Since its advent, it has been widely used in ophthalmology [2,3], dermatology [4], and angiography [5]. Its popularity is attributed to high imaging resolution and non-destructive nature. To have a better axial resolution, one needs to exploit an ultra-broadband light source. However, the dispersion, which is due to the frequency-dependent refractive index of a dispersive material, increases with the increased optical source bandwidth, causing fringe signal chirping and point spread function (PSF) broadening.

Various strategies have been used to achieve an optimal axial resolution for high-resolution OCT applications by dispersion compensation either via hardware or software-based methods. Hardware dispersion compensation includes inserting optical materials [6] or using a grating-based phase control delay line [7], both of which are capable of correcting some degree of dispersions. However, due to high cost, cumbersomeness, and the lack of interchangeability of hardware implementations, both have limited usages. Software dispersion compensation tends to be a favored solution, since it is low-cost, provides continuous tunability, and can be easily applied to other OCT platforms. Some numerical methods that directly measure the dispersion component, have been reported by Singh et al. [8] and Attendu et al. [9], where two mirror measurements at conjugate positions of the zero-delay line are used. Other numerical methods are to find the dispersion mismatch to be compensated, a phase term $e^{-i\Delta\Phi}$, multiplied with the dispersed cross-spectral density function resulting in dispersion-removed interferograms. Cense et al. [10] proposed a method to obtain the phase function $\Delta\Phi$ by polynomial fitting and obtaining the non-linear orders of the unwrapped phase of the coherence function of a well-reflecting reference point. Using mirror measurements on the sample arm at a range of different optical path differences, Uribe-Patarroyo et al. [11] performed the simultaneous k-linearization and dispersion compensation based on the fact that the correct k-mapping function is achievable by obtaining the depth-independent chirp of the dispersive fringe. Wojtkowski et al. [12] proposed an iterative procedure to attain the phase function by optimizing the image sharpness as the objective function, which compensates for second and higher-order dispersions. Furthermore, Yasuno et al. [13] used the information entropy of the image as the optimization metric. Most recently, Hillmann et al. [14] reported that Short-Time Fourier Transform (STFT) with small enough windows can be used to extract signal peak shifts by cross-correlating the sub-bandwidth reconstructed images, after which the linear part of the phase function was approximated by integrating the peak shifts $\Delta z(k)$ over $k$. Ni et al. [15] further attempted to minimize the information entropy of the spectral centroid image as the sharpness metric in the iterative optimization, where centroids of the depth-resolved spectra obtained by STFTs of fringes were calculated for each pixel corresponding to the original image. Note that some of the above-mentioned software methods are used for both the systematic and sample dispersion compensations, which could be computationally intensive and challenging for real-time image processing. While in this manuscript, since the sample dispersion was found to be negligible, we are solely targeting the systematic dispersion compensation, which can be pre-computed and calibrated only once for an SD-OCT system.

In this paper, we proposed an alternative numerical method using Time-Frequency Analysis (TFA) technique with iterative optimization for the dispersion compensation in an SD-OCT by using only one single mirror fringe. Specifically, STFT was performed on the spectral interferogram of a single mirror reflection, which resulted in a 2-D depth-wavenumber plot (TFA domain). The purpose of the automated optimization process is to minimize the wavenumber-dependent ridge variance in the TFA domain, including the variation of mirror peak position shifts in z-space after STFT and PSFs broadening, both of which are caused by the dispersion. The proposed method is simple and robust, using only one spectral interferogram, and can reduce the fringe chirping and the PSF broadening. Image resolution improvement has been demonstrated when the dispersion was compensated for measurements of the mirror and detailed biological micro-structures in the iridocorneal angle of human cadaver eyes.

## 2. Theory

*2.1 Dispersion in SD-OCT*

The interference fringes, after background subtraction, are given by Eq. (1) [16], where $k$ is the angular wavenumber, $I(k)$ is detector current, $\rho$ is the responsivity of the detector, $S(k)$ is the power spectrum of the light source, $R_R$ and $R_{Sn}$ are reflectivities in reference and sample arms, $2z$ is the optical path difference between the two arms, $\Delta\Phi(k)$ is the frequency-dependent dispersion correction term.

$$I(k) = \frac{\rho}{2} Re\left\{ S(k) \sum_{n=1}^{N} \sqrt{R_R R_{Sn}} e^{i[2k \cdot z + \Delta\Phi(k)]} \right\} \quad (1)$$

The phase function $\Delta\Phi(k)$ can be expanded as a Tayler series around the central wavenumber $k_0$:

$$\Delta\Phi(k) = \Delta\Phi(k_0) + \left.\frac{\partial \Delta\Phi(k)}{\partial k}\right|_{k_0} (k-k_0) + \frac{1}{2} \cdot \left.\frac{\partial^2 \Delta\Phi(k)}{\partial k^2}\right|_{k_0} (k-k_0)^2 + \ldots + \frac{1}{n!} \cdot \left.\frac{\partial^n \Delta\Phi(k)}{\partial k^n}\right|_{k_0} (k-k_0)^n \quad (2)$$

Typically it is sufficient to compensate the second and third-order dispersion compensations with the phase correction term as shown in Eq. (3) [12], where $a_2$ and $a_3$ are the second and third-order dispersion compensation coefficients, $k_0$ is the central angular wavenumber. Once $\Delta\Phi(k)$ is obtained, the Fourier transform of the dispersive fringe multiplied with $e^{-i\Delta\Phi(k)}$ produces a depth profile with improved axial resolution.

$$\Delta\Phi(k) = -a_2(k-k_0)^2 - a_3(k-k_0)^3 \quad (3)$$

*2.2 Time-frequency analysis (TFA)*

Time-Frequency Analyses (TFAs) are a group of techniques that represent both the time and frequency domains simultaneously, which is particularly useful for detecting non-stationary signals [17]. The most common form is the STFT. Briefly, STFT performs a size-fixed, localized segmental windowed Fourier transform, repeatedly sliding through a signal over time, after which it takes the summation of the series of windowed Fourier transforms. In practice, STFT of a resampled spectral interferogram produces a two-dimensional graphic representation of the energy distribution in the wavenumber and depth directions. While it is simple, window size selection is of extreme importance when STFT is applied to the spectral interferogram because there is a trade-off between spatial and spectral resolutions in those two directions. A narrower spectral window size results in good spectral resolution but poor spatial resolution, while a longer spectral window size will improve the spatial resolution in z with poor spectral resolution. Since we aimed to favor a better spatial resolution to ensure the accuracy in peak position determination in z-space, we chose a large spectral window length with 1024 points. Fig. 1 illustrates how STFT works and can be potentially used for targeting the systematic dispersion mismatch using a mirror measurement. The dispersion was exhibited by the sub-bandwidth reconstructed PSFs' shift and broadening, although the main cause of the latter comes from the reduced spectral bandwidth by windowing.

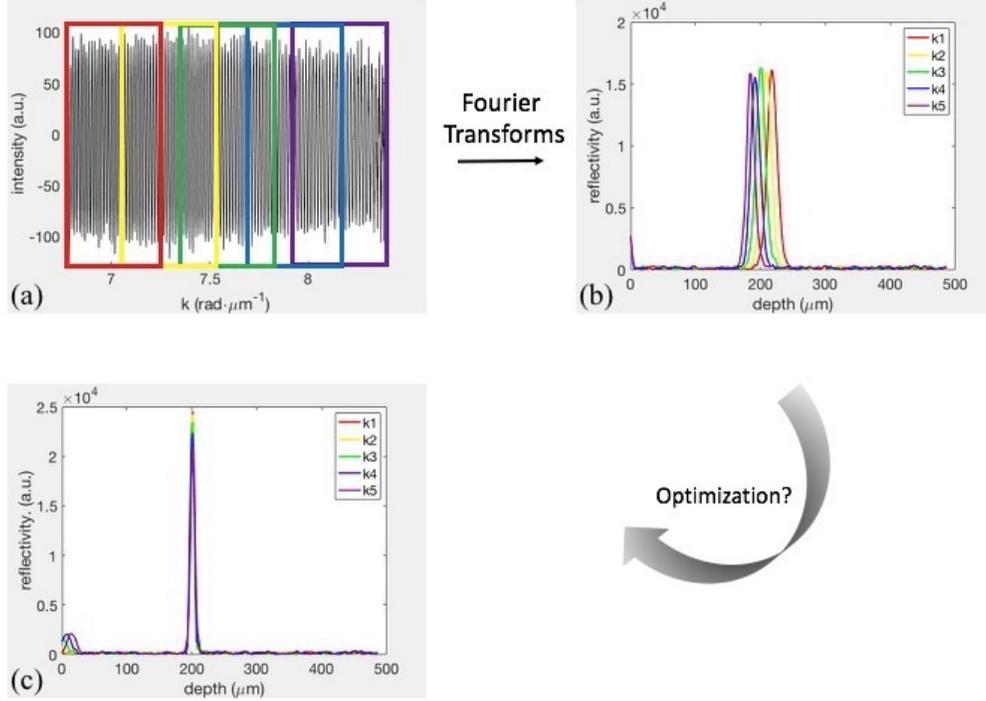

**Fig. 1.** Illustration of how STFT works on a typical resampled spectral interference signal of a mirror positioned at the sample arm. For illustration, (a) a complete fringe signal was partially captured into 5 overlapping segments by a sliding window centered at *k(i)*, where *i*=1,2,…,5. (a) Each segmented fringe was then processed via conventional reconstructions into depth profiles (b), highlighted with different colors corresponding to the window color in (a). (b) With unmatched dispersion in the system, the windowed bandwidth reconstructed PSFs are shifted and broadened accordingly. Thus, to correct the systematic dispersion mismatch, a method is needed to align all of these shifted peaks to the correct depth position as well as to sharpen the PSFs (c).

*2.3 Automatic systematic dispersion compensation algorithm using TFA*

To start the dispersion compensation term calculation, the resampled spectral interferogram, which has an equal-spacing distribution of the wavenumbers among N=2048 pixels, is input into an iterative loop that optimizes a user-defined objective function $V(a_2, a_3)$ as defined in Eq. (4). First, STFT was performed on the spectral interferogram under optimization, which resulted in a 2-D depth-wavenumber plot. The detailed MATLAB document for STFT usage can be found at https://www.mathworks.com/help/signal/ref/stft.html. The ridge was then calculated by finding the maximum depth value at each wavenumber being evaluated. The size of the wavenumbers $K_{eval}$ evaluated by STFT is determined by Eq. (5), where M is the length of the sliding window and L is the overlap length between any two consecutive sliding windows, and the ⌊ ⌋ symbols denote the floor function. The k-dependent ridge variance, caused by dispersion, was then minimized at the dispersion compensation coefficients $a_2$ or $a_3$, where $z^i_{a_2,a_3}$ is the calculated depth at $i^{th}$ wavenumber, and $\bar{z}_{a_2,a_3}$ is the average of depths across $K_{eval}$ equally spaced wavenumbers.

$$V(a_2, a_3) = \frac{\sum_i^{K_{eval}-1}\left(z^i_{a_2,a_3} - \bar{z}_{a_2,a_3}\right)^2}{K_{eval} - 1} \quad (4)$$

$$K_{eval} = \left\lfloor \frac{N-L}{M-L} \right\rfloor \quad (5)$$

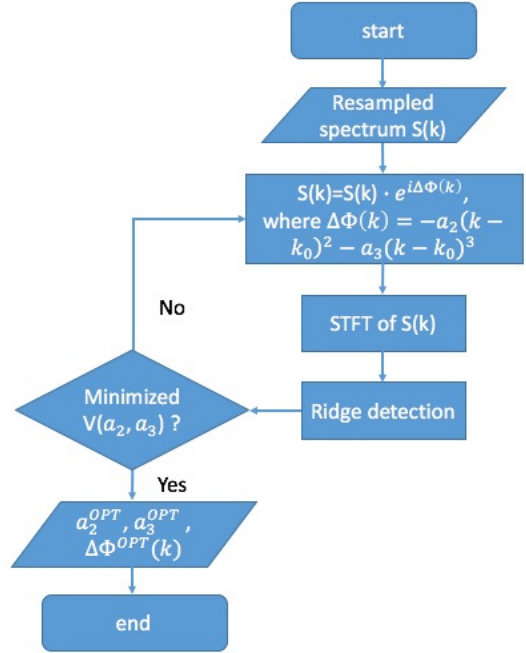

**Fig. 2.** Flowchart of our proposed SD-OCT dispersion compensation algorithm. The objective function V(a₂, a₃) denotes ridge variance, where $a_2$ and $a_3$ are variables of second and third-order dispersion compensation coefficients under optimization. STFT: Short-Time Fourier Transform; OPT: optimized.

The automated system outputs the optimized dispersion compensation coefficients $a_2^{OPT}$, $a_3^{OPT}$, and dispersion compensation term $\Delta\Phi^{OPT}(k)$, as illustrated in Fig. 2. $\Delta\Phi^{OPT}(k)$ is an N-pixel vector stored in local and is used for future dispersion compensation during routine OCT image reconstruction. Like the method described in Ref. [12], this procedure can be done for any higher-order dispersion compensations. However, since our system manifests unnoticeable third or higher-order dispersions, we calibrated the dispersion mismatch only at the second order.

### 3. Materials and Methods

*3.1 OCT system*

A schematic diagram of the custom-built SD-OCT imaging system is shown in Fig. 3. The broadband laser light source (SuperLum Ltd., Ireland) has a bandwidth of $\Delta\lambda = 165\ nm$ at 3 $dB$ spectrum width centered at 850 $nm$. The source light passing into a single-mode fiber coupler (Thorlabs, Inc.; Newton, NJ) is equally separated into two parts, directing to the reference arm and sample arm, respectively. A pair of galvanometric scanning mirrors (Cambridge Technology Inc., Bedford, MA) was implemented which enables simultaneously horizontal (X-axis) and vertical (Y-axis) scanning of the iridocorneal angle of the eye. To match the optical path lengths in the two arms, a motor-controlled mirror was used in the reference arm. A dispersion compensator (Thorlabs, Inc., Newton, NJ) was used in the reference arm to compensate for the objective lens (Thorlabs, Inc., Newton, NJ). The two lights interfere with each other at the coupler and are then detected by a spectrometer (Cobra-S 800, Wasatch Photonics, NC), which comprises a diffraction grating and a high-speed 12-bit Complementary Metal-Oxide Semiconductor (CMOS) line-scan camera.

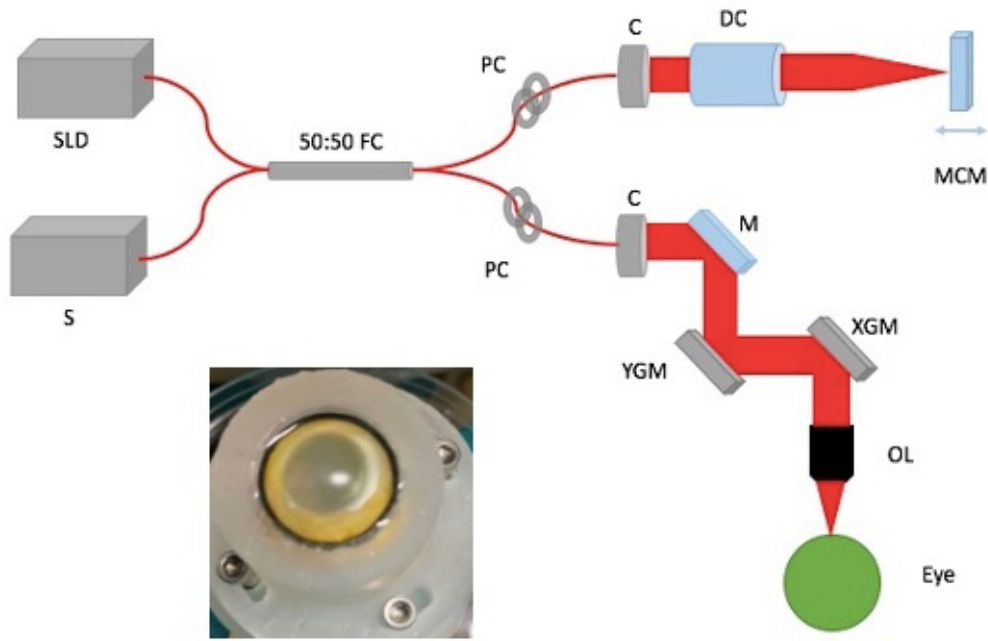

**Fig. 3.** Schematic diagram of the SD-OCT system. A systematic optical path length mismatch is corrected by using a movable mirror motor-controlled in the reference arm. The mirror or anterior segment of the cadaver eye placed on an eye holder is in the sample arm, as shown in the close-up for the cadaver eye setup. SLD: superluminescent diodes; S: spectrometer; FC: fiber coupler; PC: polarization controller; C: collimator; DC: dispersion compensator; MCM: motor-controlled mirror; M: mirror; YGM: Y-axis galvanometer mirror; XGM: X-axis galvanometer mirror; OL: objective lens.

The line-scan camera has 2048 pixels and runs rate at 20 *kHz*. Processed images contain 1024 × 1000 pixels per frame, and display images at a rate of 10 frames per second. The average axial resolution across 2-*mm* depths and theoretical lateral resolution is 2.7 *μm* and 8.2 *μm* in air, respectively. The OCT imaging system has a sensitivity of ~110 *dB* and is capable of imaging the eye to a depth of approximately 1.5 *mm* with ~4.0 *mW* incident light power on the sample.

### 3.2 Sample preparation

The human cadaver eyes for OCT imaging were obtained from San Diego Eye Bank within 24-hour postmortem. The eyes were carefully dissected to keep the trabecular meshwork (TM) intact, while removing other components, such as the posterior segment, iris, lens, ciliary body, and uveal tissues. The prepared eyes were then stored in the Optisol corneal storage medium (Bausch and Lomb, Rochester NY) and refrigerated at 4 °C. Before imaging, the enucleated eye was mounted on a customized eye holder as illustrated in Fig. 2, perfused with Dulbecco's modified Eagle's medium (DMEM) containing 5 μg/mL amphotericin, and 100 μg/mL streptomycin (MilliporeSigma, MO), and placed in an incubator (Sheldon Manufacturing, Inc., Cornelius OR) at 37 °C, 5% $CO_2$, and 90% humidity for 30 *min*. The eye was then placed on a movable mechanical stage with 5 degrees of freedom, i.e., X, Y, Z translations, rotation, and tilting under the OCT imaging system. The OCT beam was focused on the limbus to image the iridocorneal angle. The cornea surface was kept moist to prevent dehydration during imaging.

### 3.3 Methods

To test the feasibility of the proposed dispersion compensation algorithm, a physical dispersion block (LSM04DC, Thorlabs, Inc., Newton, NJ) was deliberately introduced into the reference arm in the OCT device. A mirror was placed at the focus on the sample arm while the delay line was adjusted in the reference arm such that the PSF was located at 200 $\mu m$ from zero-delay. Ten spectral interferograms were taken at this location and put into the algorithm independently to obtain the dispersion compensation vectors. Using these correction vectors, we tested the effectiveness and robustness of the algorithm by shifting PSFs at different locations as well as applying the same term to human cadaver eye data.

In performing STFT, a 1024-pixel-sized Hanning window with 99% overlapping was applied to the 2048-pixel sized spectral interferogram totaling 94 sliding spectral windows in this work. The Simplex search method ("fminsearch" function with default stopping criteria in MATLAB R2020a, MathWorks, Inc., Natick, Massachusetts) was used to minimize the ridge variance.

## 4. Results

*4.1 Systematic dispersion compensation using mirror data*

The TFA plot provides direct access to the two-dimensional representation of the step-by-step signal processing of a spectral interferogram of a single reflection of a mirror, which illustrates how the dispersion changes over the entire image processing. The energy distribution is gradually changing as shown from Fig. 4 (a-e). The energy distribution band is downward sloping and locally concentrated at some $k$ numbers after background subtraction, downward sloping and more uniformly distributed after normalization. The chirping issue was largely solved by the resampling step, but it still remained observable ridge variances. To avoid unnecessary spectral leakage effects, a Hanning window was used to suppress the fringe signals at two ends. Little ridge variance was observed after the proposed dispersion compensation.

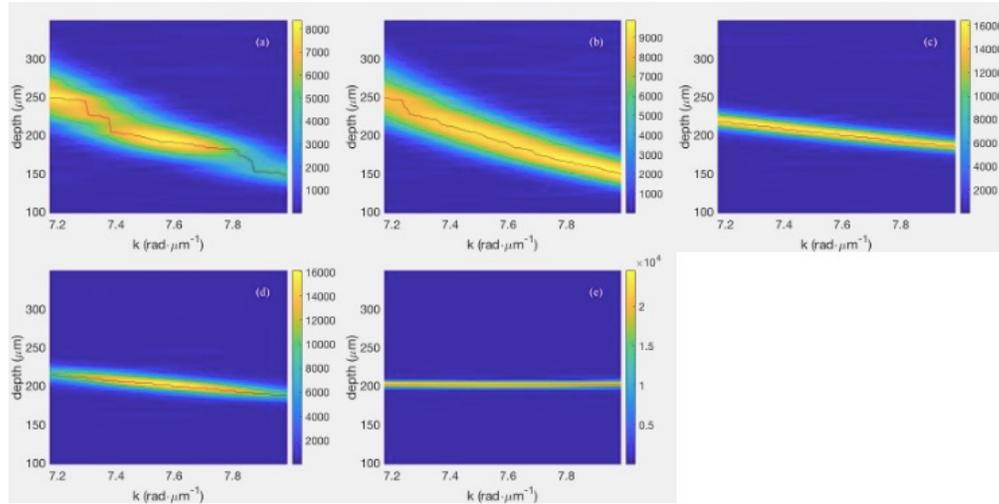

**Fig. 4.** STFT of a mirror's spectral interferogram with (a) background subtraction, (b) normalization to reference spectrum, (c) resampling, (d) windowing, and (d) the proposed dispersion compensation. The color bar indicates the amplitude of the STFT. Note that the first 50 rows of depths were deliberately removed to ensure reliable ridge detection, which corresponds to the high-intensity signals from direct current component residuals or induced from dispersion compensation. The red lines indicate the detected ridges of the energy distribution maps.

Using the proposed method discussed in Sec. 2.3, we obtained the systematic dispersion compensation term in Fig. 5 (a). It showed a quadratic curve, with a larger amount in the lower and higher frequency bands relative to the central working frequency. To avoid false ridge detections from the strong direct current component, spectral interferograms of the mirror at the focus of the sample arm were collected while shifting the PSF at 200 $\mu m$. Ten interferograms were randomly selected, calculated independently by the proposed algorithm to test the robustness of the algorithm. Fig. 5 (b) shows the reconstructed PSFs from the original, resampled, and the proposed method, respectively, with the PSFs measured at the depth of 200 $\mu m$.

The dispersion compensation vector, found to be depth-independent and once obtained, can compensate for any other measurements. Fig. 5 (c-d) summarizes the axial resolution and sensitivity roll-off performance of the imaging system over a range of 2 *mm* in air. Due to dispersion, depth reconstruction of the uncorrected spectral interferogram suffers poor axial resolutions of 11.0 to 23.5 *μm* across varying depths, with an average value of 21.2 *μm* and a standard deviation of 3.7 *μm*. With STFT dispersion compensation, however, the axial resolutions were considerably improved and range from 2.2 to 3.3 *μm*, with an average value of 2.7 *μm* and a standard deviation of 0.3 *μm*. In comparison, manual dispersion compensation was done by finely adjusting the dispersion coefficients, i.e., tuning knobs $a_2$ and $a_3$ according to Eq. (3) to have the perceptibly narrowest PSF, which provided an axial resolution of 3.0 ± 0.3 *μm*.

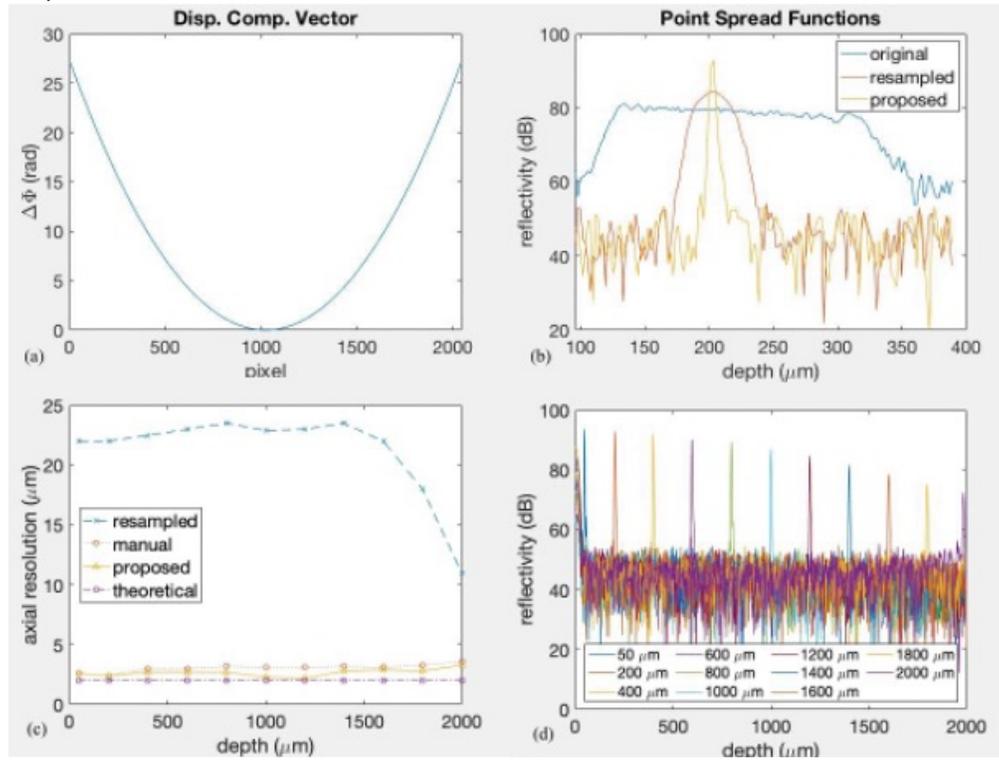

**Fig. 5.** Dispersion compensation vector and system performances. (a) The dispersion compensation vector was obtained from the proposed method. (b) The PSFs of the original, resampled, and STFT-iterative corrected spectra were logarithm scaled. (c) Measured OCT axial resolution in air differs with varying axial positions obtained by different methods - no dispersion compensation, with dispersion compensation using manual adjustment, or the proposed STFT and iterative optimization method, and theoretical values, and (d) the sensitivity roll-off performance of the calibrated SD-OCT system by the proposed method.

*4.2 Imaging for iridocorneal angle in human cadaver eyes*

Visualization of the iridocorneal angle details provides ophthalmologists/researchers a direct view of the ocular outflow structures, such as the trabecular meshwork (TM), Schlemm's canal (SC), collector channels (CCs), and vessels in the corneoscleral limbus region [18]. The TM and the nearby SC are both ring structures and are sometimes hard to find on the OCT images given its deep-seated nature and shadowing effect caused by the overlying blood vessels [19]. CCs [20] are smaller opening structures typically 30 *μm* in diameter with large variation and occasionally appeared connecting the outer wall of the SC and episcleral veins. These structures, affected by some physiological or pathological changes, are directly or indirectly responsible for reduced aqueous outflow in human eyes, leading to elevated eye pressure and glaucoma development. To restore the natural ocular aqueous outflow, one can drain into some part of the TM tissue [18] by different methods, e.g, medical, surgical, and laser treatments [21,22].

The current gold standard for angle imaging associated with glaucoma is gonioscopy [23], which, can view the TM surface through cornea but it's extremely subjective and operator-dependent. For the potential glaucoma treatment application, we implemented the OCT imaging system specifically for the iridocorneal angle in human eyes. A *2mm×1.5mm* area in human cadaver eyes was shown in Fig. 6. To remove the speckle noise, 11 sequential images at the same location were taken to obtain an averaged image. The dispersion compensated images were sharper and showed more clearly the anatomical details in the iridocorneal angle. For instance, SC, CCs, and adjacent vessels were clearly visualized on the dispersion compensated images while barely visible on uncorrected images.

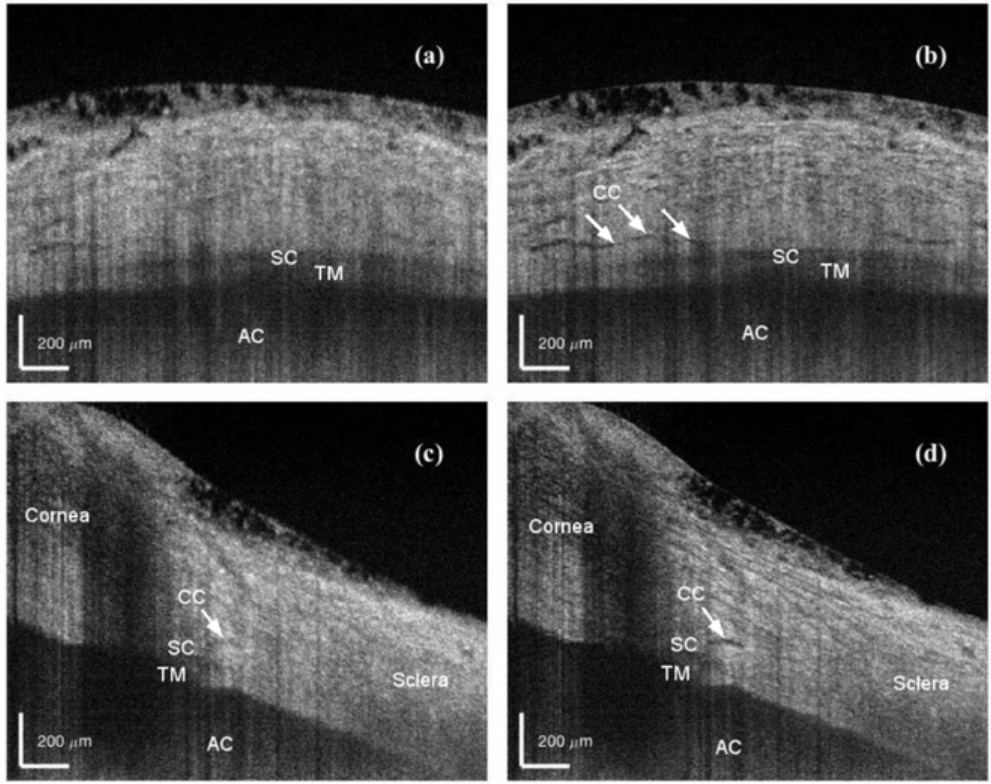

**Fig. 6.** Typical averaged horizontal (a, b) and vertical scanning (c, d) OCT images of the iridocorneal angle of human cadaver eyes with a field of view $2mm \times 1.5mm$, compared between (a, c) without and (b, d) with the proposed dispersion compensation algorithm. The beam probe was arranged approximately perpendicular to the limbus to ensure the best visibility of TM, SC, and surrounding microstructures. SC, CCs, and adjacent vessels were found on the dispersion compensated images while some of them were not visible or blurred on the uncorrected images. Scale bar is 200 *μm*. AC: anterior chamber; TM: trabecular meshwork; Schlemm's canal: SC; CC: collector channel.

## 5. Discussion

Optical dispersion remains a challenge in the development of an OCT system, especially when using light sources with broad bandwidth, such as the 165 *nm* spectral bandwidth used in our study. In addition, the dispersion is difficult to detect in the spectral interferogram domain. To address these challenges, we developed a new TFA automated iterative framework that extracted the systematic dispersion compensation by optimizing the energy distribution of a dispersive spectrum from a reflection off a mirror. The mirror data showed that dispersion induced discontinuity or disturbances and k-dependent fluctuation of ridges in the TFA domain. This variance was minimized via an iterative procedure together with STFT. An intuitive observation is that, if we take an integral along the horizontal axis to eliminate the k-dependence in the TFA domain, the FWHM of the resultant 1-D PSF can be greatly reduced after ridge variance minimization (Fig. 1).

Previous studies have attempted to use STFT and/or iterative optimization methods for correcting both systematic and sample dispersion compensations [14,15], which might be challenging for real-time processing. Alternatively, by simply scanning a mirror at the sample arm in this work, the systematic dispersion was implemented and pre-calibrated for the SD-OCT. Unlike using cross-correlation to find the peak shifts [14], we used a simple maximum search method to detect the locations of ridges, followed by an iterative optimization procedure for ridge variation minimization. Ni et al.'s work [15] focused on the optimization of the sample centroid image' information entropy, as a further refinement on the researches of Wojtkowski et al. [12] and Yasuno et al. [13] towards optimizing an image's contrast and entropy, respectively. Although these studies have improved the image qualities, extra computation time is needed. In some clinical cases, this might be a burden especially for real-time practice, such as monitoring before and during glaucoma treatment applications. Considering these and the negligible sample dispersion compensation in our iridocorneal angle application, we, therefore, attempted to calibrate only the systematic dispersion compensation. Our algorithm is simple, robust, and automatic. The dispersion compensation terms obtained from the 10 spectral interferograms of a mirror positioned in the sample arm have only a -0.46% coefficient of variation (Table 1). The proposed method was also successfully tested for manually adjusted mirror measurements, which were thought to have been dispersion corrected, demonstrating its extraction capability with an imperceptibly small amount of dispersion (Table 2). Lastly, to our best knowledge, the algorithm requires the least amount of data and the easiest involvement of the system for calibration. This could be potentially served as a universal initial calibration step to any other SD-OCT scanners for systematic dispersion compensation.

**Table 1. Statistics of the Extracted $a_2$ Values from 10 Mirror Fringes with Adding a Dispersion Compensating Block**

| | |
|---|---|
| extracted $a_2$ values ($\times 10^{-11} m^2 \cdot rad^{-1}$) | -4.098, -4.098, -4.098, -4.098, -4.098, -4.144, -4.133, -4.121, -4.133, -4.098 |
| mean ($\times 10^{-11} m^2 \cdot rad^{-1}$) | -4.118 |
| coefficient of variation | -0.0046 |

**Table 2. Statistics of the Extracted $a_2$ Values from 10 Manually Matched Mirror Fringes**

| | |
|---|---|
| extracted $a_2$ values ($\times 10^{-12} m^2 \cdot rad^{-1}$) | -4.657, -4.657, -4.657, -4.657, -4.657, -3.725, -3.725, -3.725, -3.725, -3.725 |
| mean ($\times 10^{-12} m^2 \cdot rad^{-1}$) | -4.191 |
| coefficient of variation | -0.1171 |

The resolution and imaging capability of the angle details were improved with the proposed method using the STFT iterative technique. The axial resolutions were consistently superior to the manual adjustment method across varying depths within 2 *mm* in air, the latter of which is based on the same dispersion compensation formula (Eq. (2)) but requires a subjective and time-consuming tuning process. Importantly, the dispersion compensation algorithm enabled high-fidelity visualization of CCs adjacent to the SC, which was lost on the uncompensated images. Furthermore, simultaneous horizontal and vertical imaging can assist to access angle details more efficiently, which is crucial before and during glaucoma surgeries. For instance, in Fig. 6(b), the horizontal scanning allows ophthalmologists to have a different view of the continuous structures as compared to the cross-sectioned structures in the vertical scanning in Fig. 6(d). This would be beneficial for the treatment planning of the surgical volumes in our ongoing laser studies in human cadaver eyes, because laser drilling through TM that closes to the opened SC/CCs might have a positive effect on intraocular pressure reduction. Given the importance of better image visualization and characterization of the ocular outflow structures in the iridocorneal angle for glaucoma surgery, future studies include image quality improvement of the TM/SC/CC areas, as well as the studies of OCT image-guided laser surgery for the treatment of glaucoma.

In conclusion, we presented a new method for the systematic dispersion compensation for an SD-OCT, for imaging the iridocorneal angle of human cadaver eyes. The dispersion compensation term can be calculated with an automatic iterative procedure that minimizes the k-dependent ridge variance via optimization of energy redistribution of a mirror's spectral interferogram in the TFA domain using STFT. Lastly, we demonstrate the feasibility of the proposed method for systematic dispersion compensation in an SD-OCT by evaluating both the mirror and angle imaging of human cadaver eyes.

**Funding**. National Institutes of Health (5R01 EY030304-02); Research to Prevent Blindness, Inc. (RPB-203478).

**Disclosures**. The authors declare no conflicts of interest.

**Data availability.** Data underlying the results presented in this paper are not publicly available at this time but may be obtained from the authors upon reasonable request.